\documentclass{iopjournal}
\usepackage{braket}
\usepackage{amssymb}
\usepackage{tabularx}
\usepackage{amsmath}
\usepackage{etoolbox}
\usepackage{ragged2e}
\usepackage{cite}
\begin{document}
\justifying 
\articletype{Paper} %	 e.g. Paper, Letter, Topical Review...

\title{Characterizing quantum synchronization in the van der Pol oscillator via tomogram and photon correlation}

%\author{Kingshuk Adhikary\footnote{Authors contributed equally}$^{\dagger, \top1}$, K. M. Athira\footnotemark[1]$^{\star,\|2}$, M. Rohith$^{\star,\|3}$}

\author{Kingshuk Adhikary\footnote[2]{Authors contributed equally}$^{,\,1,\, 2,\,\intercal}$, \orcid{0000-0003-0034-9889}, K. M. Athira\footnotemark[2]$^{,\, 3,\,4,\,\ddagger}$\orcid{0009-0007-7547-6678} and M. Rohith$^{3,\,4,*}$\orcid{0000-0002-5722-8319}}

\affil{$^1$School of Physical Sciences, Indian Association for the Cultivation of Science, Jadavpur, Kolkata 700032, India}

\affil{$^2$Optics and Quantum Information Group, The Institute of Mathematical Sciences, CIT Campus, Taramani, Chennai 600113, India.}

\affil{$^3$P. G. \& Research Department of Physics, Government Arts and Science College Kozhikode, University of Calicut, Kozhikode 673 018, India}

\affil{$^4$Quantum Systems Lab, Department of Physics, Government College Malappuram, University of Calicut, Munduparamba 676509, India}

\email{$^*$rohithmanayil@gcmalappuram.ac.in (Corresponding Author), $^\intercal$kingshuka@imsc.res.in, and\\$^\ddagger$athirakmkakkadath@gmail.com}


\begin{abstract}
\justifying 
Scalable methods for detecting and quantifying the nonclassical nature of a quantum state in noisy environments are challenging due to a complex relationship between noise and quantum coherence. In particular, identifying experimentally accessible signatures of synchronization in such regimes remains an open problem. By leveraging promising experimental implementation, we underpin what possible direct measures of nonclassicality are available. This work outlines accessing quantum synchronization (QS) in the steady state of a driven quantum van der Pol oscillator (vdPo) using two distinct figures of merit: (i) the nonclassical area $\delta$ and (ii) the second-order correlation function $g^{(2)}(0)$, both of which are viable in experimental architectures. The nonclassical area quantifier based on homodyne tomography allows us to assess the nonclassical nature of the vdPo state directly from the tomogram without requiring full state reconstruction or Wigner function negativity. Within a well-defined parameter regime of drive strength and detuning, both $\delta$ and $g^{(2)}(0)$ exhibit pronounced signatures of synchronization that complements the phase coherence between the drive and the vdPo. We derive an analytical expression for the steady state density matrix and the corresponding tomogram of the system, valid for arbitrary strengths of the harmonic drive. Analysis of the quantum tomogram uncovers clear phase locking behaviour, enabling the identification of the synchronization region (Arnold tongue) directly in terms of $g^{(2)}(0)$ and $\delta$. Furthermore, the behaviour of $g^{(2)}(0)$ provides a statistical perspective that reinforces the tomographic signatures of QS.  By analyzing the interplay between the aforementioned metrics, our findings indicate a scalable and experimentally relevant framework for characterizing QS in the driven vdPo.
\end{abstract}

\section{Introduction}
Synchronization \cite{Pikovsky2001,Strogatz2004} is a fundamental phenomenon wherein interacting systems adjust their rhythms to exhibit coherent dynamics. This behaviour is ubiquitous in classical nonlinear systems, from pendulum clocks to neural networks, and is typically driven by the interplay between nonlinearity and nonequilibrium features, leading to gain and loss processes in coupled systems. Beyond its classical roots, synchronization has also been identified in quantum mechanical systems \cite{Ramakrishnan2023, RAMAKRISHNAN2024129396, Godonou_2024}, where it has garnered significant interest for its potential to deepen our understanding of complex dynamics and to enable technological innovations in areas such as quantum metrology \cite{Gaurav2025}, communication \cite{Jakub2021,Sun2024}, and computation \cite{Mahlow2024,Martin2020}. Thus in the quantum regime, quantum synchronization (QS) \cite{Hush2015,Weiss2016,Amitai2017,Witthaut2017,Roulet2018,Manzano2013,Laskar2020,Lee2013,Lee2014,Walter2014,Sonar2018,Martin2020} offers insight into how collective coherence arises from the continuous competition between coherent driving and dissipation. These studies demonstrate the importance of nonlinear dissipation \cite{Lee2013,Sonar2018,Walter2014} in stabilizing synchronized states while also revealing noise-induced synchronization and synchronization blockade \cite{Martin2020, Laskar2020}. Along these lines, driven-dissipative synchronization results from the interaction of coherent dynamics and environmental coupling, which is frequently manifested as phase coherence \cite{Lee2014} and entanglement \cite{Manzano2013} buildup. In this context, quantum fluctuations \cite{Weiss2016} can qualitatively modify classical synchronization \cite{Witthaut2017} behaviour, leading to regimes absent in their classical counterparts. However, QS is not merely a scaled-down version of its classical counterpart, it also involves distinctly quantum features, such as quantum fluctuations, coherence, and entanglement \cite{Roulet2018b}, which make its characterization a nontrivial task. The phenomenon of QS has been studied extensively in various engineered quantum systems \cite{Heinrich2011, Kreinberg2019,Ludwig2013, Lee2013,English2015,Li2025} and quantum analogs of classical oscillators \cite{Lee2014, Mari2013}. Among these van der Pol oscillator (vdPo) \cite{VanderPol1926} stands out as a prototypical model to explore QS. It provides a minimal yet effective framework where gain and loss mechanisms, characteristic of open quantum systems \cite{Walter2014,Sonar2018,Mok2020,Jaseem2020}, are balanced with nonlinearity to produce synchronized steady states. 

The driven vdPo is a powerful model to study the quantum-classical boundary which in the classical limit show limit cycles and synchronization due to external driving while in quantum domain retains these characteristics but influenced by the quantum features. Several works \cite{Lee2013,Lee2014,Walter2014,Cabot2021,Kato2023,Wachtler2023,Li2025} have analyzed the behaviour of a single driven quantum vdPo and its synchronization to an external frequency of the drive. These studies show that a single driven quantum vdPo can exhibit phase locking to an external drive, closely resembling classical synchronization in the appropriate limit. At the same time, they reveal distinctly quantum features, where fluctuations and energy quantization modify the stability and sharpness of synchronization. Such systems provide a versatile platform to explore transitions between classical and quantum synchronization regimes under controlled driving and dissipation. Early studies demonstrated that a driven quantum vdPo can exhibit phase locking and synchronization plateaus despite quantum noise \cite{Lee2013,Lee2014}, while frequency entrainment and locking were identified through power spectral analysis \cite{Walter2014}. Classical-quantum links in synchronization have been identified in the exact dynamics of a quantum vdPo \cite{Witthaut2017}. The emergence of metastable phases were observed for a squeezed-driven \cite{Sonar2018} quantum vdPo where system intermittently switches between different synchronized states and is unique to quantum regime \cite{Cabot2021}. With Kerr nonlinearity \cite{Yurke1986}, the quantum asymptotic phase revealed signature of quantum torus synchronization where multiple frequencies lock rather than a limit cycle in the deep quantum regime \cite{Kato2023}. Topologically coupled vdPo network showed edge-state synchronization robust to noise and disorder \cite{Wachtler2023}. Recently, an experimental implementation of a driven quantum vdPo using $^{40}$Ca$^+$ atom inside a Paul trap has been proposed, which can be a breakthrough in bringing QS to trapped ion platforms \cite{Li2025}.

In quantum systems, synchronization is often visualized via Arnold tongue \cite{Arnold1961} structures, with various measures proposed \cite{Mari2013,Walter2014,Ameri2015,Roulet2018,Jaseem2020} to quantify it. Tools like the Wigner function \cite{Wigner1932} or Husimi function \cite{Husimi1940} and frequency entrainment are commonly used but require reconstruction from measured observables, which is error-prone and computationally demanding for large Hilbert spaces. Recent advances suggest that quantum tomograms \cite{Bellini2012,Rohith2015} can directly capture state features without requiring full state reconstruction. In this work, we develop a tomographic framework to analyze the dynamics of a driven quantum vdPo and characterize synchronization behaviour. As part of this, we use the {\em nonclassical area} ($\delta$)  \cite{Rohith2023} as a synchronization-sensitive quantifier, constructed from the phase dependent quadrature variance extracted from the quantum tomogram. Nonclassical area  measures deviations from the vacuum or coherent state benchmark, which represents the most classical like quantum states and saturates the minimum uncertainty bound. Thus it quantifies the effective area projected by a tomogram on the tomographic plane referenced against that of a vacuum state. Owing to its tomographic definition, $\delta$ serves as a useful tool for probing synchronization in the quantum regime as phase locking between the drive and the oscillator manifests through characteristic angular modulations of quadrature fluctuations. Moreover, quantum state tomography \cite{Cramer2010,Christandl2012,Stricker2022} enables the practical estimation of $\delta$ by measuring the rotated quadrature operator and constructing the corresponding tomogram as a positive probability distribution \cite{Gilson1982}. In this context, an important challenge is to identify synchronization signatures that remain robust in the presence of noise and are accessible without full state reconstruction. Building on these foundations, our primary objective is to establish experimentally feasible and scalable methods for detecting and quantifying nonclassicality and synchronization in a vdPo. Consequently, the framework developed here bridges the gap between theoretical characterization and experimental detection of synchronization in open quantum systems.

In this manuscript, we propose a new measure of QS that builds upon insight from quantum tomograms to quantum optics. The synchronization of the vdPo to the external drive is an excellent framework to investigate universal synchronization behaviour. To measure synchronization in a driven quantum vdPo, we employ the nonclassical area $\delta$ as a diagnostic figure of merit. In particular, we investigate the emergence of nonclassical Arnold tongue structures, paralleling the classical Arnold tongue in the quantum vdPo, through both the nonclassical area and the equal time second-order correlation function \cite{Fox2006}. In this work, we study the synchronization of quantum vdPo into two extreme limits, the classical (characterized by $\kappa_2/\kappa_1=0$) and deep quantum regimes ($\kappa_2/\kappa_1\ge10$), as per the classification in Ref. \cite{Mok2020}. We benchmark the QS measured from both figures of merit for the driven vdPo under evidence of the Arnold tongue shape. Finally, in the landscape of QS measured in the deep quantum limit ($\kappa_2/\kappa_1\rightarrow\infty$), we demonstrate the construction of the quantum state in the quadrature plane with the tomogram and in phase space with the Wigner distribution. Furthermore, to achieve direct experimental access, we reformulate the master equation of driven vdPo in terms of the quantum tomogram, avoiding full state reconstruction.

% \textcolor{red}{We analyze the temporal evolution of quantum tomograms under varying driving strengths and quantify nonclassicality using the nonclassical area metric.} \textcolor{red}{We also identify signatures of limit-cycle oscillations in the steady-state quantum tomograms. In each case, we derive the steady-state density matrix in the deep quantum regime and construct the corresponding tomograms to assess nonclassical features.} 

The content of this paper is structured as follows: In section \ref{sec2}, we develop a model of driven-dissipative vdPo. In section \ref{sec2a}, we discuss the assessment of QS on the prospects of the tomographic representation of quantum states and introduce the nonclassical area as an indicator of nonclassicality. In addition, another figure of merit, two particle correlation, will be used to quantitatively compare the synchronization behaviour of quantum vdPo. We display our results regarding the appearance of QS in section \ref{sec3}. In section \ref{sec4}, we examine how quantum tomograms evolve at steady state within the framework of an Arnold tongue in the deep quantum regime. Finally, section \ref{sec5} summarizes the key findings of this work with the prospects of an experimental feasibility of QS. We described the computational methodology of the steady state density matrix in appendix \ref{algo}.

\section{Theoretical framework and figure of merits}\label{sec2}
A quantum vdPo serves as an ideal platform for studying QS with an external drive on account of unavoidable environmental degrees of freedom. The dynamics of the whole dissipative system in the rotating frame of the external drive are governed by the following master equation (in units of $\hbar$) for the density matrix $\rho$, which has the standard form (in the rotating frame with the external drive)\cite{Lee2014,Walter2014}:
\begin{equation}
\frac{d{\rho}}{d t} = -i[ \Delta a^\dagger a + F(a + a^\dagger), \rho] + \kappa_1 \mathcal{D}[ a^\dagger ]\rho + \kappa_2 \mathcal{D}[ a^2 ]\rho,
\label{mes}
\end{equation}
where $a$ $(a^{\dagger})$ is the annihilation (creation) operator of the single mode oscillator, $\Delta$ is the detuning of the drive’s frequency $\omega_d$ from the natural frequency of the oscillator $\omega_0$, and $F$ denotes the drive strength. The Lindblad dissipator is defined as $\mathcal{D}[\mathcal{O}]\rho = (2\mathcal{O} \rho \mathcal{O}^{\dagger} - \mathcal{O}^{\dagger} \mathcal{O} \rho - \rho \mathcal{O}^{\dagger} \mathcal{O})/2$, which accounts for environment-induced dissipation, while $\kappa_1$ and $\kappa_2$ represent the rates of linear pumping and nonlinear damping, respectively. \(\mathcal{D}[ a^\dagger ]\rho \) assemble single excitation into the system while \(\mathcal{D}[ a^2 ]\rho\) removes a pair of excitation from the system. The competition between linear pumping and nonlinear damping gives rise to self-sustained oscillations (limit cycle) with stable amplitude, and is crucial for the emergence of synchronization behaviour in the quantum regime. A natural direction to follow at this point is to consider characterization of synchronization in the long-time regime, where the system reaches a steady state determined by the interplay of coherent driving and dissipation. The numerical simulations of the master Eq. (\ref{mes}) are done at the steady state $(t\rightarrow\infty)$ in QuTip \cite{Johansson2013} to obtain the steady state density matrix $\rho^{ss}=\rho(t\rightarrow\infty)$. Notably, all system parameters are scaled by the linear damping rate $\kappa_1$, unless otherwise specified.

Strictly speaking, we do not explore the synchronization behaviour across different dynamical regimes; instead, we focus on two extreme limiting cases by tuning $\kappa_2$ in Eq. (\ref{mes}). These regimes correspond to the classical limit, realized by setting \(\kappa_2=0\) and the deep quantum limit, achieved by taking \(\kappa_2 \to \infty\). In the classical limit, the absence of nonlinear damping eliminates two-photon loss, and the system dynamics are governed primarily by linear gain and external driving, thereby suppressing quantum features and yielding behaviour close to that of a classical oscillator.  Conversely, in the deep quantum limit, the nonlinear damping term dominates thus playing a major role in characterizing the discrete level structure of vdPo. Physically, large $\kappa_2$ corresponds to extremely strong two-photon loss, which rapidly de-excites higher photon-number states at rates proportional to $\kappa_2 n(n-1)$. As a result, population in Fock states with $n\geq 2$ is strongly suppressed. Here \(\kappa_2\) effectively restricts the system’s evolution to the lowest Fock states, mostly to the subspace spanned by the vacuum ($|0\rangle$) and single-photon ($|1\rangle$) states , by strongly suppressing higher photon-number excitations.

\subsection{Measure of quantum synchronization}\label{sec2a}

Unlike the synchronization in classical systems, the QS cannot be addressed directly through conventional means, since the fluctuations of the phase space variables strictly maintain the Heisenberg principle. A main open question is about the possibility of measuring QS in the presence of a different kind of environment coupling that causes dephasing rather than dissipation. In this study, we address this question by discussing the assessment of two distinct figures of merit: (i) quantum tomogram and (ii) quantum correlation, in a framework of the quantum vdPo. The first one encoded peculiar features in quantum measurement that are directly related to experimentally accessible probability distributions \cite{Gilson1982,Zambrano2025} (e.g., quadrature measurements in quantum optics) and reveal nonclassical features like squeezing, superposition, and entanglement. In addition, the tomographic representation exploits universal applicability across systems and is neither basis dependent nor linked to any particular operator basis (like photon number or position). The latter figure of merit, $g^{(2)}(0)$ \cite{Fox2006, Adhikary2021}, is directly measurable in experiments \cite{ko2025} and provides an invasive signature on novel many-body phenomena, including quantum phase transitions that account for fundamental quantum fluctuations in promising physical setups such as Bose-Hubbard, Dicke model, circuit-QED, etc. By analyzing both of these figures of merit, we can gain more profound insights into the underlying processes that facilitate synchronization in our quantum network. To the best of our knowledge, the development of a theory-experimental bridge with the aforementioned figure of merit in the context of QS has yet to be discussed.

\subsubsection{Tomographic framework}

Having detailed the driven-dissipative quantum vdPo we plan to investigate in our work, let us now highlight the figure of merit for quantifying synchronization in tomographic pictures. This picture offers a practical and robust alternative to probe synchronization particularly for experimental implementations. Here we characterize the QS in the vdPo through its quantum tomogram. The rotated quadrature operators form a quorum of observables sufficient to reconstruct any single mode quantum state. The single mode rotated quadrature operator is given by
\begin{equation}
     \mathbb{X}_{\theta}=\frac{1}{\sqrt{2}}( a e^{-i\theta}+ a^\dagger e^{i\theta}),
\end{equation}
where \(\theta\in[0,2\pi]\) is the rotation angle in phase space that determines which field quadrature is being measured and $[ a, \mathbb{X}_{\theta}]=e^{i\theta}/\sqrt{2}$. The quantum tomogram is the probability distribution of measurement outcomes for \(\mathbb{X}_{\theta}\). For  quantum state described by the density matrix \(\rho\) the tomogram is defined as \cite{Lvovsky2009,Bellini2012}
\begin{equation}
    \omega(X_{\theta}, \theta)=\langle X_{\theta}, \theta|\rho|X_{\theta}, \theta\rangle.
    \label{ome}
\end{equation}
Here, $\mathbb{X}_{\theta}\left|X_{\theta}, \theta\right\rangle=X_{\theta}\left|X_{\theta}, \theta\right\rangle$ with 
\begin{eqnarray}
    |X_{\theta}, \theta\rangle&=&\frac{1}{\pi^{1/4}}\exp\left[-\frac{X^2_{\theta}}{2}-\frac{1}{2}e^{i2\theta}a^{\dagger 2}+\sqrt{2}e^{i\theta}a^{\dagger}X_{\theta}\right]|0\rangle\nonumber\\
    &=&\sum_{nm}\mathcal{A}_{nm}|n+2m\rangle,
    \label{eq4}
\end{eqnarray}
where $\mathcal{A}_{nm}=\frac{2^{\frac{n}{2}-m}\sqrt{(n+2m)!}}{\pi^{\frac{1}{4}}n!m!}X^n_\theta e^{[i(n+2m)\theta+im\pi-\frac{X^2_{\theta}}{2}]}$.

Nonclassicality quantifier derived directly from the tomogram offers a practical framework to investigate QS. For a better understanding of the nonclassical area of a single mode field, it is defined as \cite{Rohith2023}
\begin{equation}
\delta\left(\rho\right)=\int_{0}^{2\pi} d\theta \,\Delta X_\theta-\sqrt{2}\,\pi,\label{Nonclassical_Area} 
\end{equation} 
where \(\Delta X_{\theta}\) is the standard deviation in the measurement of the rotated quadrature operator $\mathbb{X}_{\theta}$. Defined through quadrature fluctuations, $\delta = 0$ for classical states and $\delta \ne 0$ for genuinely nonclassical states, with higher values indicating stronger nonclassical character. Unlike indicators that may also respond to classical statistical mixtures, \(\delta\) effectively filters out classicality and captures intrinsic quantum features.

\subsubsection{Quantum correlation}

Another commonly used indicator of nonclassicality, especially in photon statistics, is the equal time second-order correlation function, $g^{(2)}(0)$, which characterizes the nature of correlations between two quanta (photons, excitations, etc.) emitted at the same time. At the steady state, for a single mode vdPo, $g^{(2)}(0)$ is defined as \cite{Fox2006},
\begin{equation}
    g^{(2)}(0)=\frac{\langle a^\dagger a^\dagger aa\rangle_{\rho^{ss}}}{\langle a^\dagger a \rangle_{\rho^{ss}}^2}.  \label{second_order_correlation}
\end{equation}
It is a measure of quantum degree of coherence encompassing information about photon correlations in the field. This insight reveals the non-universal connection between $g^{(2)}(0)$ and QS. It strongly depends on the physical system, parameter regime, and how synchronization can reach the fundamental limit imposed by the laws of quantum mechanics. This dependence highlights the importance of tailoring experimental setups to explore the nuances of QS.

% \begin{table*}[t]
% \renewcommand{\arraystretch}{1.9} % Adjust the value as needed
% \begin{center}
%     \begin{tabularx}{0.7\textwidth} { 
%    >{\raggedright\arraybackslash}X 
%   | >{\centering}X 
%   | >{\raggedright\arraybackslash}X  }
%  \hline\hline
% \textbf{Synchronization regime} & $\mathbf{ g^{(2)}(0)}$ & \textbf{Statistical indicator} \\
%  \hline
%  Unsynchronized, random & $<1$ & Irregular, anti-bunching \\
%  Synchronized, single-quantum regime & $\approx1$ & Regular, phase-locked, random \\
%  Synchronized, collective bursts  & $>1$  &  Highly regular, bunching   \\
% \hline\hline
% \end{tabularx}
% \end{center}
% \caption{Synchronization performance of the quantum vdPo with statistical fingerprint. The results are bound by either maximum or minimum synchronization. This classification indicates that the efficiency of the synchronization process can vary significantly depending on the specific conditions and parameters involved in our study. }
% \end{table*}

\begin{table*}[t]
\renewcommand{\arraystretch}{1.4}
\centering

\begin{tabularx}{1\textwidth}{
>{\raggedright\arraybackslash}X 
| >{\centering\arraybackslash}X 
| >{\raggedright\arraybackslash}X
}
\hline\hline
\textbf{Synchronization regime} & $g^{(2)}(0)$ & \textbf{Statistical indicator} \\
\hline
Unsynchronized, random & $<1$ & Irregular, anti-bunching \\
Synchronized, single-quantum regime & $\approx 1$ & Regular, phase-locked, random \\
Synchronized, collective bursts & $>1$ & Highly regular, bunching \\
\hline\hline
\end{tabularx}

\caption{Synchronization performance of the quantum vdPo with statistical fingerprint. The results are bound by either maximum or minimum synchronization. This classification indicates that the efficiency of the synchronization process can vary significantly depending on the specific conditions and parameters involved in our study. }
\end{table*}

\section{Nonclassical manifestations of synchronization}\label{sec3}

Our investigation of synchronization measure employs the nonclassical area ($\delta$) formalism as a powerful tool for visualizing the Arnold tongue with nonclassicality. In particular, we examine the behaviour of $\delta$ as a function of the driving strength $F$ and detuning $\Delta$, across two distinct dynamical regimes defined by the nonlinear damping parameter $\kappa_2$. We consider a comparative analysis of the system dynamics in both deep quantum and classical regimes, with particular emphasis on the phase locking phenomenon. The Arnold tongue is a characteristic structure in the parameter space of driven nonlinear systems, often used to illustrate regions of synchronization. It delineates the range of parameters, typically drive strength and detuning, over which the system exhibits phase locking.
\begin{figure}[!h]
\centering
\includegraphics[scale=0.45]{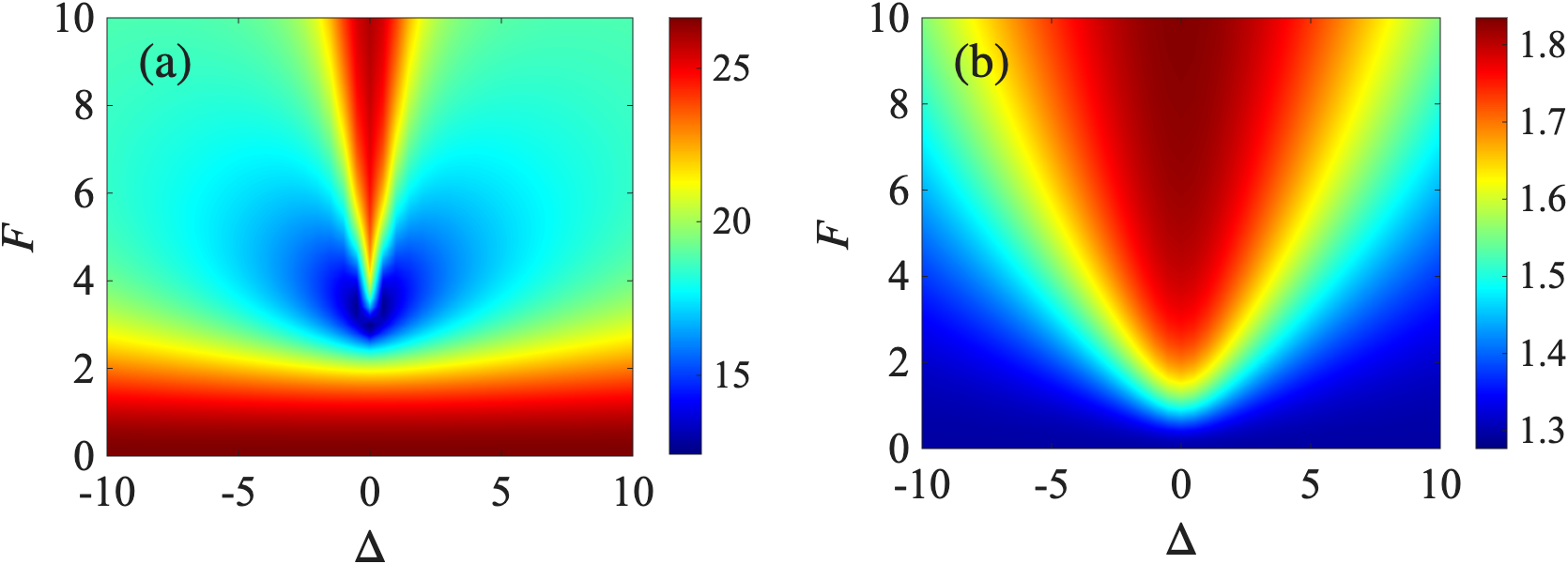}\\
\vspace{0.2in}
\includegraphics[scale=0.4]{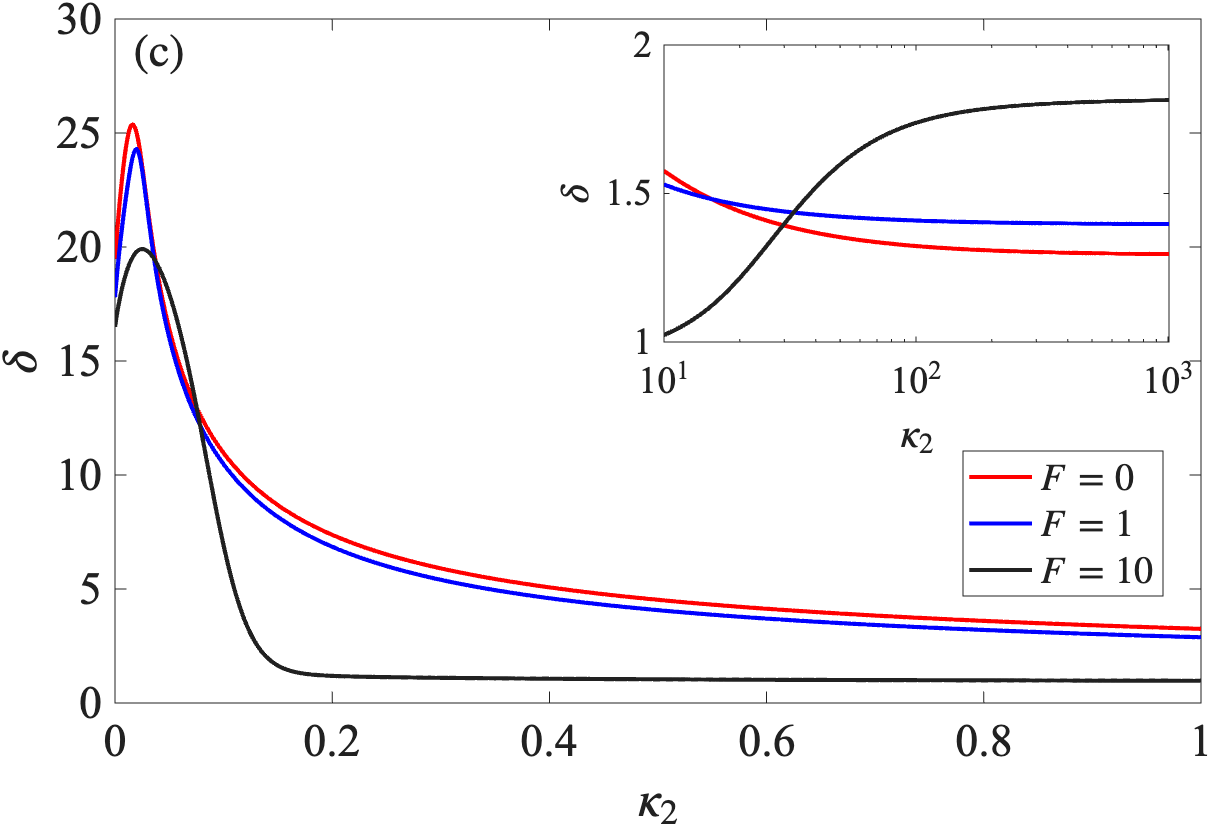}
\caption{(Color online) Different behaviours of the phase locking measure $\delta$ at the steady state. The nonclassical area $\delta$ measures synchronization at the steady state of an externally driven vdPo. In panels (a) and (b), red (blue) indicates maximum (minimum) synchronization as measured by $\delta$. The damping ratio is set to (a) $\kappa_2=0$, (b) $\kappa_2=10^3$. Inside the tongue: strong phase locking, collective dynamics $(\delta>>1)$. Outside the tongue: Weak or no phase locking, less collective emission $(\delta>1)$. The lower panel (c) represents the synchronization measure across various regimes of the quantum vdPo, as mentioned in Reference \cite{Mok2020} in Table 1. The different drive strengths $F$, indicated in the legend, are selected from the top panels, implying a scenario of undriven to driven quantum vdPo, with $\Delta=2$. Increasing the driving strength is seen to produce qualitative nonclassicality up to the quantum regime ($\kappa_2=1$). The inset illustrates that the behaviour of nonclassicality in the deep quantum regime ($\kappa_2 >> 1$) results in synchronization similar to that observed in the classical limit.}
\label{arn_tongue}
\end{figure}

Figure~\ref{arn_tongue} displays the Arnold tongue shape of the nonclassical area \( \delta \) given in Eq. (\ref{Nonclassical_Area}) as functions of $F$ and $\Delta$. Panels (a) and (b) correspond to the classical limit (\( \kappa_2 = 0 \)) and the deep quantum limit (\( \kappa_2 = 10^3 \)), respectively. Since \( \delta \) is unbounded by any upper limit, an increase in \( \delta \) indicates enhancement in nonclassicality. In both panels, red (blue) regions indicate maximal (minimal) values of \( \delta \), representing strong (weak) nonclassicality and hence stronger (weaker) synchronization signatures. In our study, the driving strength $F$ is considered up to an order of 10, consistent with values already realized in experiments \cite{sudler2024,Li2025}. In the classical regime [figure~\ref{arn_tongue}(a)], \( \delta \) exhibits a pronounced dependence on both \( F \) and \( \Delta \), forming a distinct Arnold tongue structure centered around \( \Delta = 0 \). This tongue-shaped feature marks the region in parameter space where the oscillator locks to the drive frequency, indicating enhanced synchronization. In classical limit $\kappa_2=0$, the weak drive strength $F<1$ always capture the maximum synchronization (red) but strong drive only measure the synchronization around the resonance. However, in $(F-\Delta)$ plane, a certain region (blue) indicates absence of any synchronization. For higher driving the nonclassical area reaches values as high as \( \delta \sim 26 \) in the vicinity of resonance.

In the deep quantum limit, characterized by strong nonlinear damping ($\kappa_2=10^3$), the dynamics of the quantum vdPo are confined to the lowest Fock levels, specifically $|0\rangle$ and $|1\rangle$. As shown in figure~\ref{arn_tongue}(b), the nonclassical area remains significantly reduced with values saturating around $\delta \sim 1.8$. Despite this reduction in nonclassicality, the system still exhibits a well-defined synchronization region. The Arnold tongue structure, though less sharp, becomes more spread out, forming a broad triangular region centered at resonance (\( \Delta = 0 \)). This indicates a smooth crossover from the unsynchronized to synchronized regime as the driving strength \( F \) increases. Unlike the classical case where synchronization onset is abrupt and accompanied by a sharp increase in \( \delta \), the deep quantum regime displays a more gradual phase locking behaviour with relatively modest changes in the nonclassical area. The key conclusion drawn from panels ~\ref{arn_tongue}(a) and (b) is that the Arnold tongue illustrates the synchronization region with phase locking in parameter space by plotting the nonclassical area measure against the driving strength $F$ and detuning $\Delta$.

Figure ~\ref{arn_tongue}(c) helps us understand the synchronization as well as nonclassical measure smoothly from the classical to deep quantum regime, whereas the top panels~\ref{arn_tongue}(a, b) are in two extreme regimes of the quantum vdPo in the $(F-\Delta)$ plane. The observation is that the classical region inherently shows strong synchronization with $\delta>1$; conversely, in the quantum limit, the system becomes less synchronized. As a result, the classical limit $\kappa_2=0$ captures strong synchronization with the higher nonclassicality, whereas the deep quantum regime is exactly opposite in it. Importantly, in all these cases, our results show that there is typically an inverse relationship between the nonclassical area and QS in the vdPo.

\begin{figure}[h]
    \centering
\includegraphics[width=6.0in,height=1.8in]{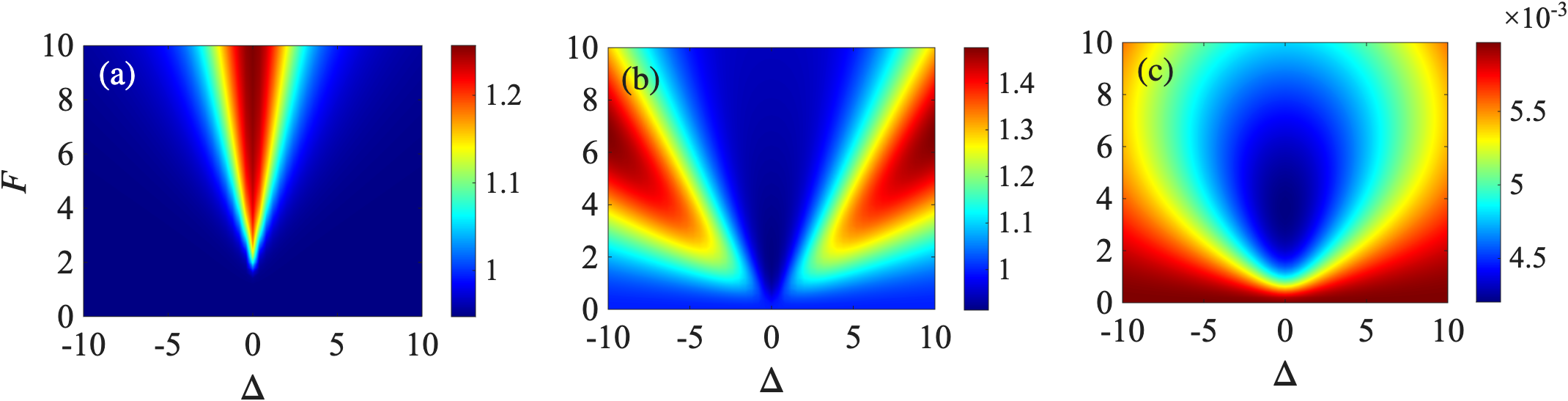}
\caption{(color online). At steady state, the equal time second-order correlation $g^{(2)}(0)$ exhibits phase locking as a function of driving strength $F$ and detuning $\Delta$. In panels (a, b), the red (blue) region indicates the maximum (minimum) synchronization measurement. Inside the tongue: strong phase locking indicates bunching with collective dynamics $(g^{(2)}(0)>1)$. Outside the tongue: Weak or no phase locking indicates antibunching with nonclassicality due to $(g^{(2)}(0)<1)$. In panel (c), the absence of Arnold's tongue shape, indicated by $g^{(2)}(0) \rightarrow 0$, highlights a strikingly diminishing correlation with strong nonclassicality. The damping ratio is set to (a) $\kappa_2=0$, (b) $\kappa_2=1$, and (c) $\kappa_2=10^3$.}
\label{arn_tongue_g2}
\end{figure}
To gain a more comprehensive understanding of the synchronization measure, we study the quantum correlation at steady state, which is expected to affect phase locking significantly. In this regard, in Figure \ref{arn_tongue_g2}, the $g^{(2)}(0)$ function is drawn as functions of $F$ and $\Delta$ for different values of the $\kappa_2$. In Figure \ref{arn_tongue_g2}(a), the vdPo is strongly synchronized in the classical limit $\kappa_2=0$ with the drive; it can exhibit phase lock, collective oscillations, leading to bursty or bunched emission ($g^{(2)}(0) > 1$). Here, enhanced synchronization is linked with bunching. It is observed that for large drive strength or detuning, synchronization typically increases. The intriguing feature in the quantum regime $(\kappa_2=1)$ is that mirror symmetry of the Arnold tongues with respect to $(\Delta = 0)$ arises from the system’s symmetry: synchronization can occur for drive frequencies both above and below the natural frequency by the same amount (in terms of $|\Delta|$), as shown in figure~\ref{arn_tongue_g2}(b). On the other hand, it is observed that at the deep quantum regime $\kappa_2=10^3$, synchronization is not captured, resulting in $g^{(2)}(0)\rightarrow0$ due to the strong quantum fluctuation in the phase of vdPo, as shown in figure~\ref{arn_tongue_g2}(c).

Overall, the synchronization characteristics from $g^{(2)}(0)$ lead to the emergence of phase locking and its corresponding photon statistics, as summarized in Table 1. We would like to emphasize that antibunched, coherent, and bunched regimes indirectly link synchronization. The stark signature of $g^{(2)}(0)$ emphasizes that the sensitivity of synchronization is dependent on how well the operating regime of quantum vdPo's is controlled through careful selection of system parameters. This is a distinctive signature of how QS manifests in nonlinear, driven-dissipative quantum systems like the vdPo.

\section{Analytics of tomogram and phase coherence}\label{sec4}

In the Fock basis, the master equation (\ref{mes}) for the driven vdPo can be expressed in the matrix-element form as
\begin{eqnarray}
    \frac{d}{dt}\rho_{m,n}&=&-i\Delta(m-n)\rho_{m,n}\nonumber\\&&-iF\left(\sqrt{m}\rho_{m-1,n}+\sqrt{m+1}\rho_{m+1,n}-\sqrt{n}\rho_{m,n-1}-\sqrt{n+1}\rho_{m,n+1}\right)\nonumber\\&&+\frac{\kappa_1}{2}\left(2\sqrt{mn}\rho_{m-1,n-1}-(m+n+2)\rho_{m,n}\right)\nonumber\\&&+\frac{\kappa_2}{2}\left(2\sqrt{(m+1)(m+2)(n+1)(n+2)}\rho_{m+2,n+2}-(m^2+n^2-m-n)\rho_{m,n}\right)
    \label{mn_mastereqn}
\end{eqnarray}

At steady state, where $\frac{d}{dt} \rho_{m,n}=0$, the density-matrix elements $\rho_{m,n}$ are calculated numerically in QuTiP \cite{Johansson2013} and also analytically for the small system.

In the quadrature plane $(X_\theta-\theta)$, the master equation~(\ref{mes}) in the eigenbasis $\left|X_{\theta}, \theta\right\rangle$ of the  single mode rotated quadrature operator $\mathbb{X_\theta}$ can be written as, 
\begin{eqnarray}   
    \frac{d}{d t}\rho_{p+2q;n+2m}=-i\Delta\Big((p+2q)\mathcal{A}^*_{pq}-(n+2m)\mathcal{A}_{nm}\Big)\nonumber\\-iF\mathcal{A}^*_{pq}\Big(\sqrt{p+2q}\rho_{p+2q-1;n+2m}+\sqrt{p+2q+1}\rho_{p+2q+1;n+2m}\Big)\nonumber\\+iF\mathcal{A}_{nm}\Big(\sqrt{n+2m}\rho_{p+2q;n+2m-1}+\sqrt{n+2m+1}\rho_{p+2q;n+2m+1}\Big)\nonumber\\+\frac{\kappa_1}{2}\mathcal{A}_{nm}\mathcal{A}^*_{pq}\Big(2\sqrt{(n+2m)(p+2q)}\rho_{p+2q-1;n+2m-1}-(n+p+2m+2q+2)\rho_{p+2q;n+2m}\Big)\nonumber\\+\frac{\kappa_2}{2}\mathcal{A}_{nm}\mathcal{A}^*_{pq}\bigg(2\sqrt{(n+2m+1)(n+2m+2)(p+2q+1)(p+2q+2)}\rho_{p+2q+2;n+2m+2}-\nonumber\\\Big((p+2q)(p+2q-1)-(n+2m)(n+2m-1)\Big)\rho_{p+2q;n+2m}\bigg).
\label{meqt}
\end{eqnarray}
This tomogram-based master equation~(\ref{meqt}) results in a significant corollary: it allows for a potential direct measurement of the synchronization with the nonclassical area $\delta$ of the quantum vdPo. It is important to note that this corollary is numerous and far-reaching after transforming the master equation~(\ref{mes}) apart from vdPo. For driving strengths up to $F\leq10$, which are experimentally accessible \cite{Mok2020}, the one-photon coherent drive primarily couples two consecutive energy eigenstates of the undriven vdPo.

The existence of coherence terms in $\rho^{ss}$ depends on the driving strength $F$, while their magnitude is governed by the nonlinear damping rate $\kappa_{2}$. Consequently, we neglect higher-order coherence terms in $\rho^{ss}$ beyond the deep quantum regime, and the relevant Hilbert space can be truncated to $|0\rangle$, $|1\rangle$, and $|2\rangle$, with coherences only between $|0\rangle$ and $|1\rangle$ for $F \leq 10$. Based on this, we adopt the following ansatz for the steady state density matrix $\rho^{ss}$ of the driven vdPo in the deep quantum regime:
\begin{equation}
    \rho^{ss}= 
\begin{pmatrix}
\rho_{00} & \rho_{01} & 0\\
\rho_{10} & \rho_{11} & 0\\
0 & 0 & \rho_{22}
\end{pmatrix},
\label{steadystate_driven}
\end{equation}
where the nonvanishing off-diagonal elements $\rho_{01}$ and $\rho_{10}=\rho_{01}^*$ explicitly account for the coherence between the lowest-energy states $|0\rangle$ and $|1\rangle$. Imposing the steady state condition $\dot{\rho}=0$ on equation~(\ref{mn_mastereqn}), the following set of coupled linear equations:

\begin{equation}
\left.
\begin{array}{rcl}
    0 & = & -iF(\rho_{10} - \rho_{01}) - \kappa_1\rho_{00} + 2\kappa_2\rho_{22} \\
    0 &= & iF(\rho_{10} - \rho_{01}) + \kappa_1(\rho_{00} - 2\rho_{11}) \\
    0 & = & 2\kappa_1\rho_{11} - (3\kappa_1 + 2\kappa_2)\rho_{22} \\
    0 &= &(i\Delta - \frac{2}{3}\kappa_1)\rho_{01}- iF(\rho_{11} - \rho_{00}) \\
\end{array}
\right\}.
\quad
\end{equation}

Solving the above coupled equations gives full expressions of the steady state matrix elements:
\begin{equation}
\left.
\begin{array}{rcl}
    \rho_{00} &=&\frac{2}{R}\left(6F^2 + 4\Delta^2 + 9\kappa_1^2\right) \\
    \rho_{01} &=& \rho_{10}^* = \frac{2F}{R}\left(3i\kappa_1 - 2\Delta\right) \\
    \rho_{11} &= &\frac{1}{R}\left(12F^2 + 4\Delta^2 + 9\kappa_1^2\right) \\
    \rho_{22} &=& 1 - \frac{3}{R}\left(8F^2 + 4\Delta^2 + 9\kappa_1^2\right)
\end{array}
\right\},
\label{b2s}
\end{equation}
where the denominator $R=(12F^2+4\Delta^2+9\kappa_1^2)(3+\frac{\kappa_1}{\kappa_2})-12F^2$. In the limit $\kappa_2\rightarrow\infty$, the maximum population is bound up to $|2\rangle$ and neglects all coherences beyond the state  $|0\rangle-|1\rangle$ manifold. Without higher-order coherences, analytical calculations are much more straightforward as well as negligible in exact numerical simulations. In the absence of drive, $F=0$, and in the limit $\kappa_2\rightarrow\infty$, we are able to write the elements of equation~(\ref{steadystate_driven}) as $\rho_{10}=\rho_{22}=0$, $\rho_{00}=2/3$, and $\rho_{11}=1/3$. Correspondingly, the steady state given in equation~(\ref{steadystate_driven}) reduces to a limit cycle \cite{Lee2013,Lee2014}:
\begin{eqnarray}
    \rho^{ss}_{\kappa_2\rightarrow\infty; F=0}=\frac{2}{3} |0\rangle\langle 0|+\frac{1}{3} |1\rangle\langle 1|,
    \label{steadystate}
\end{eqnarray}
reveals a statistical mixture of the vacuum and single photon state, with no phase coherence existing between the Fock levels owing to the vanishing off-diagonal terms in equation~(\ref{steadystate}). Since higher Fock levels are suppressed by the large $\kappa_2$, the system predominantly relaxes to $|0\rangle$ while retaining a finite probability of occupying $|1\rangle$ because of the dissipative dynamics. In principle, $\rho^{ss}_{\kappa_2\rightarrow\infty}$ is therefore a mixed state with nonzero entropy and no phase information, precluding the possibility of phase synchronization in this regime. Notably, this limit cycle \cite{Lee2014} has a ring-shaped positive Wigner function in the phase space.

The full expression for the steady state tomogram $\omega(X_{\theta}, \theta)$ with the advent of density matrix elements \ref{b2s} is given by
\begin{eqnarray}
    \omega(X_{\theta}, \theta)
&=&\frac{e^{-X_\theta^2}}{2\sqrt{\pi}R}\Bigg(X_\theta^4\left(R-24F^2-12\Delta^2-27\kappa_1^2\right)\nonumber\\&&-2X_\theta^2\left(R-48F^2-20\Delta^2-45\kappa_1^2\right)\nonumber\\&&-8\sqrt{2}FX_\theta(3\kappa_1\sin{\theta}+2\Delta\cos{\theta})+\left(4\Delta^2+9\kappa_1^2+R\right)\Bigg),
\end{eqnarray}
and we can simplify the above expressions for the undriven vdPo in limit $\kappa_2\rightarrow\infty$ as
\begin{equation}
\omega(X_{\theta})_{\kappa_2\rightarrow\infty;F=0}=\frac{2}{3\sqrt{\pi}}(1+X^2_\theta)e^{-X^2_\theta}.
\label{eq12}
\end{equation}

\begin{figure}[t]
\centering
  \begin{tabular}{@{}ccc@{}}
     \hspace{-.2in}
    \includegraphics[height=2.3in, width=6in]{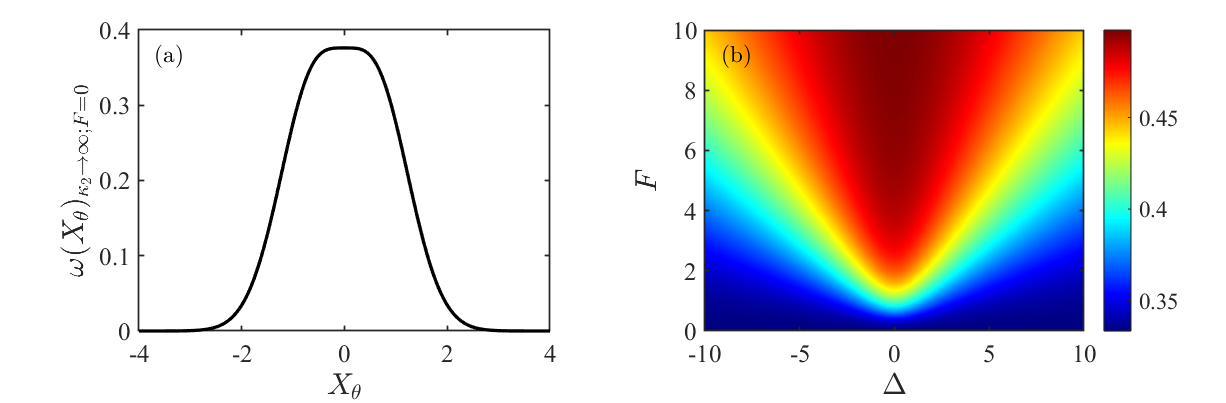} 
     \end{tabular}
 \caption{(color online). (a) Quantum tomogram against quadrature eigenvalues $X_\theta$ for an undriven quantum vdPo showing rotational symmetry, i.e., no preferred phase direction. From the synchronization perspective, that corresponds to no phase locking. The limit cycle amplitude equation~(\ref{eq13}) is shown in panel (b) at steady state.  The color scale indicates the degree of phase locking: deep-blue regions correspond to weak synchronization, while the transition to red marks enhanced synchronization-near resonance ($\Delta\approx0$) and grows with increasing drive strength.}
    \label{N_steadystate}
\end{figure}

\begin{figure}[h]
    \centering
\includegraphics[width=6in,height=1.8in]{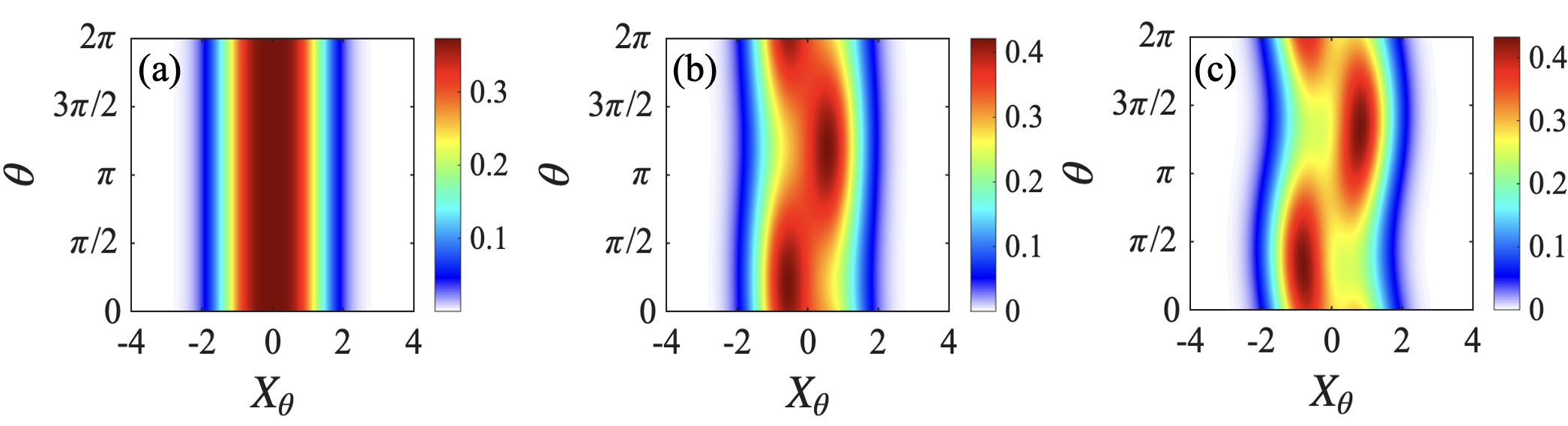}\vspace{-0.03in}
\includegraphics[width=6in,height=1.8in]{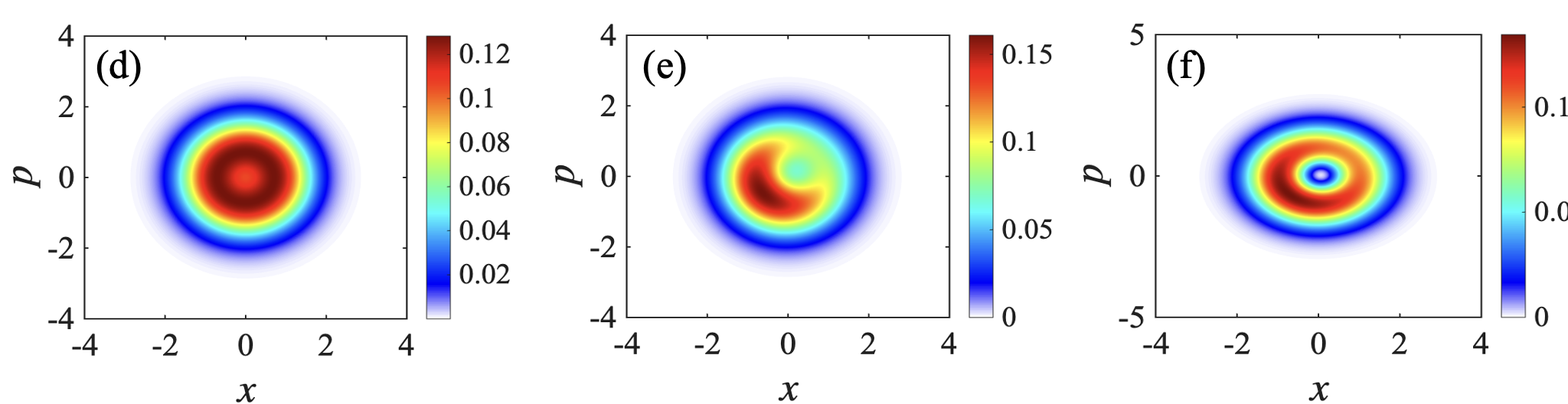}
\caption{(color online). Top row: Tomographic probability distributions $\omega(X_\theta,\theta)$ at steady state of the driven quantum vdPo in the deep quantum regime, shown for different drive strengths: (a) weak drive $F=0$, (b) intermediate drive $F=1$, and (c) strong drive $F=10$. For weak driving [panel (a)], the tomogram remains broadly distributed, indicating low phase preference and weak synchronization. At intermediate driving [panel (b)], the distribution becomes anisotropic with visible deformation, signaling the onset of partial phase locking. Under strong driving [panel (c)], the tomogram exhibits a pronounced localized shape, characteristic of enhanced phase concentration and stronger synchronization. Bottom row: Wigner functions (d-f) corresponding to the tomograms in panels (a–c). At weak drive (d), the Wigner function retains near-rotational symmetry, consistent with the absence of synchronization. For intermediate drive (e), angular modulation emerges, marking the onset of rotational symmetry breaking. Under strong drive (f), the Wigner function shows pronounced phase-space asymmetry, mirroring the tomogram’s localization and providing clear evidence of QS. Here, we choose $\Delta=2$ for all panels.}
    \label{Fig:Tomogram_Wigner}
\end{figure}

We note that the observed tomogram $\omega(X_\theta, \theta=0)$ in figure~\ref{N_steadystate} from equation~(\ref{eq12}) is a Gaussian-modulated polynomial, which is completely symmetric under rotation, as shown earlier in figure~\ref{Fig:Tomogram_Wigner}(a). This typically corresponds to vacuum or thermal states, indicating no synchronization.

Next, we discuss how the limit cycle [equation~(\ref{steadystate})] amplitude is determined by the mean excitation number, $N=\langle a^\dagger a\rangle=\rho_{11}+2\rho_{22}$ in the deep quantum limit ($\kappa_2\rightarrow\infty$). The analytical expression is subsequently calculated using elements (\ref{b2s}) as follows:
\begin{eqnarray}
N_{\kappa_2\rightarrow\infty}&=&\frac{1}{3}+\frac{4F^2}{24F^2+12\Delta^2+27\kappa_1^2}.
\label{eq13}
\end{eqnarray}
The constant term $1/3$ sets the amplitude of the limit cycle, which corresponds to the intrinsic quantum fluctuations for the undriven vdPo, as explained in equation~(\ref{steadystate}). This baseline value agrees with the well-known finding that sheds light on the classification of the operating regime characterized by $\kappa_2/\kappa_1$, as reported in Ref.~\cite{Mok2020}, confirming that the undriven quantum vdPo sustains a nonzero excitation number due to quantum noise. To understand the parameter dependence of the limit cycle, we display it in figure~(\ref{N_steadystate}), showing the Arnold tongue, which is a key signature of synchronization. The effect of drive enables slowly increasing the amplitude and prominently captures the maximum mean excitation at resonance $\Delta=0$. In essence, in the deep quantum limit, the limit cycle amplitude reflects a nontrivial interplay between quantum fluctuations, coherent drive, and dissipation.

The Wigner distribution in phase space summarizes the striking physical insight of $\omega(X_{\theta}, \theta)$ that defines the different measures of synchronization in the quadrature plane. This approach provides a comprehensive understanding of the visualization (see figure~\ref{Fig:Tomogram_Wigner}) of quantum states and their features in the perspective of QS. Figure~\ref{Fig:Tomogram_Wigner} exhibits the behaviour of the steady state quantum tomograms \ref{Fig:Tomogram_Wigner}(a)-\ref{Fig:Tomogram_Wigner}(c) and their corresponding Wigner functions \ref{Fig:Tomogram_Wigner}(d)-\ref{Fig:Tomogram_Wigner}(f) at varying driving strengths in the deep quantum limit. Let us now be more precise.

In the absence of drive ($F=0$), the tomogram \ref{Fig:Tomogram_Wigner}(a) remains essentially independent of the quadrature angle $\theta$, reflecting full angular symmetry and the absence of synchronization. In phase space, the associated Wigner function \ref{Fig:Tomogram_Wigner}(d) displays a rotationally symmetric distribution (disk shape) centered at the origin , indicating no preferred phase and thus no phase locking to the external drive. This symmetry reflects complete phase indeterminacy: the system does not prefer any particular phase. At weak drive $F=1$, the tomogram \ref{Fig:Tomogram_Wigner}(b) develops clear oscillations along $\theta$, signifying the onset of phase localization. This oscillation reflects the emergence of phase sensitivity in the quadrature outcomes due to partial phase locking to the external drive. As expected, QS induces a spontaneous symmetry breaking with weak driving in the Wigner function (see figure~\ref{Fig:Tomogram_Wigner}(e)), forcing the previously circular (rotationally symmetric) Wigner function to become deformed. This deformation indicates that the system is developing a preferred phase, the hallmark of synchronization. For strong drive $F=10$, figure~\ref{Fig:Tomogram_Wigner}(c) exhibits that the oscillations are sharper, showing that the system's quadratures are tightly locked to the external drive, which is a signature of strong QS. The corresponding Wigner function in figure~\ref{Fig:Tomogram_Wigner}(f) shows pronounced phase-space localization (banana-like shape) shifted away from the center with reduced rotational symmetry. Thus, reflecting tomograms in the phase space configuration offers a clearer understanding of synchronization by revealing the emergence of phase dependent quadrature statistics.

\begin{figure}[ht]
    \centering   
    \includegraphics[width=6in,height=2.1in]{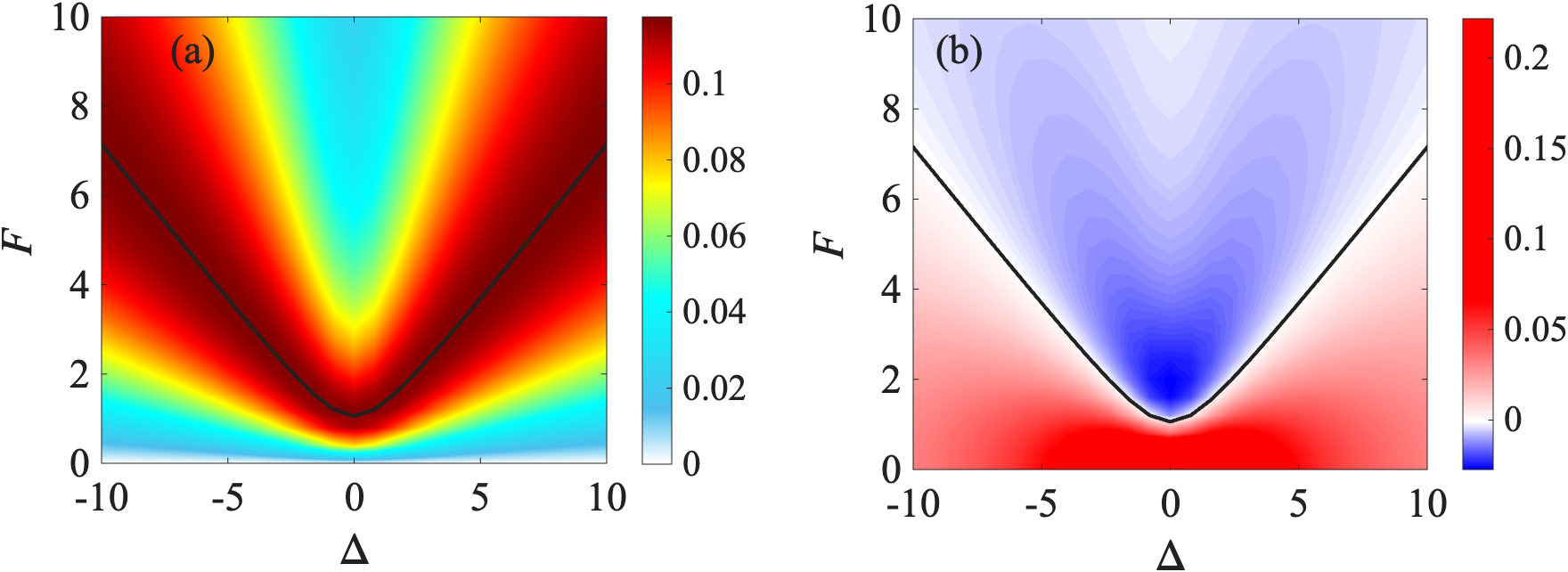}
    \caption{(color online). (a) Steady state coherence $|\rho_{01}|$ as a function of driving strength $F$ and detuning $\Delta$ for the quantum vdPo in the limit $\kappa_2\rightarrow\infty$, showing sensitivity of phase locking. Outside of the red area, the vdPo exhibits more classical (mixed) behaviour, loses quantum features, and fails to exhibit synchronization. (b) The gradient $\partial_F\mathcal{S}$ is used to locate the synchronization threshold, or the region where quantum control is most effective. By fine-tuning the parameters, the phase relationship between $|0\rangle$ and $|1\rangle$ is optimized, enabling more reliable qubit operations. In both panels, the black curve represents the critical drive $F_c$ that describes the peak of maximum coherence $|\rho_{01}|$, which inherently demonstrates a two level qubit.} 
    \label{coh}
\end{figure}

Finally, we analyze the phase coherence ($\mathcal{S}$) that encodes the relative phase between vdPo and the external drive, indicating the degree of phase locking inside the system. This analysis is crucial and describes how phase coherence can influence QS in the limit $\kappa_2\rightarrow\infty$. It
is straightforward to reach the explicit form of
the relative phase with as follows
\begin{eqnarray}
    \mathcal{S}_{\kappa_2\rightarrow\infty}&=& |\rho_{01}|\nonumber\\
               &=& \frac{2F\sqrt{4\Delta^2+9\kappa_1^2}}{24F^2+12\Delta^2+27\kappa_1^2},
\end{eqnarray}
which is easily found by solving for the steady state density matrix elements of $\rho^{ss}$, as defined in equation~(\ref{steadystate_driven}). The connection between synchronization and coherence has already been reported in literature, for example, in Refs. \cite{Mok2020, PhysRevA.109.033718}. Notably, here we address the coherence in a fully quantum mechanical way by specifically targeting the off-diagonal element, $\rho_{01}$, which only survives in the limit $\kappa_2\rightarrow\infty$ and is also restricted for the drive $F\le10$. A full characterization of the emergence of coherence $\rho_{01}$ is given in figure~\ref{coh}(a) where, by varying both $F$ and $\Delta$, we calculate the maximum coherence $\mathcal{S}(F_c)\approx0.12$ determined from the critical drive strength $F_c=\sqrt{(9\kappa_1^2+4\Delta^2)/8}$, as indicated by the black curve. This value highlights the optimization of coherence amplitude, which adequately supports the drive strength of $F \le 10$, and also reveals the maximum phase locking in limit $\kappa_2 \rightarrow \infty$. As expected, the classical regime $\kappa_2\rightarrow0$ refers to lower coherence $\mathcal{S}\approx0.006$ (maximum) due to discarding all quantum fluctuations. Since $\rho_{22}=0$, the population imbalance $|\rho_{00}-\rho_{11}|$ can be fostered to optimize the coherence $\mathcal{S}$ in parallel with $\rho_{01}$ using the oscillator state $\rho^{ss}_{\kappa_2\rightarrow\infty}$. Eventually, controlling the excitations with the drive force ($F\neq0$) allows for a distribution that is restricted to two discrete eigenbases, $|0 \rangle$ and $|1 \rangle$ of the vdPo, which is represented as a conceptual two level qubit. This distribution facilitates the mapping of a quantum vdPo into a qubit, which itself is synchronized with the external drive.

The results in Figure \ref{coh}(a) motivate us to comprehensively examine the sensitivity of coherence; we have made use of the interpretation of the gradient $\partial_F\mathcal{S}$ [see in figure~\ref{coh}(b)], which acts as an onset for the synchronization measure. In stark contrast to the qubit picture, which emerges from the coherence solely influenced by the drive $F$, we gain insights into the phase locking between the qubit and the external drive. The positive (negative) slope indicates an in phase (anti phase) relationship between $|0\rangle$ and $|1\rangle$, but the critical drive $Fc$ describes maximum coherence between them, completely absent in the generic limit cycle $\rho^{ss}_{\kappa_2\rightarrow\infty; F=0}$. The results indicate that simulating a driven vdPo as a two-level qubit in the limit $\kappa_2 \rightarrow \infty$ reveals a phase locking aspect related to synchronization measurement.

\section{Conclusions}\label{sec5}

In this work we have developed a tomographic framework to characterize synchronization in a driven quantum vdPo. The central figures of merit in this work are the nonclassical area $\delta$ and the second-order correlation function $g^{(2)}(0)$, both of which access synchronization in different regimes classified by $\kappa_2/\kappa_1$. By employing the nonclassical area as a quantifier, we identified Arnold tongue structures in the quadrature space that clearly reflect synchronization signatures in both the classical and quantum regimes. In the classical limit, the onset of synchronization is sharp, accompanied by a significant increase in nonclassicality, whereas in the deep quantum regime, the synchronization regions become broader and smoother, with the nonclassical area saturating to modest values. Interestingly, the second-order correlation function delivers statistical signatures of synchronization.

The basic theoretical contribution of this study is the explicit derivation of the steady state density matrix elements at arbitrary driving strengths in the deep quantum regime, where nonlinear damping confines the oscillator to the lowest Fock levels. This analytic solution captures how drive strength and detuning shape the populations and coherences of the steady state. We also analyzed the steady state tomograms for different driving strengths, which provided an experimentally accessible visualization of the underlying nonclassical features. Likewise, our tomogram measure is a key achievement that delineates an underlying set of quantum states in the QS landscape. Importantly, we observe symmetry breaking and phase localization in the quantum tomograms, providing a direct and novel signature of the onset of QS. This finding is complementary to the symmetry breaking in the Wigner function, linking  phase space asymmetry with measurable tomographic features of synchronization. Similarly, the nutshell of the two level qubit is crucial for phase locking as coherence emerges in the deep quantum limit. In addition, we have reformulated [equation~(\ref{meqt})] the master equation in terms of the quantum tomogram, providing direct access to synchronization signatures from experimentally measurable probability distributions. These results offer a practical framework for the robustness of nonclassical steady states in vdPos with potential relevance for future experimental and technological applications, including quantum state engineering \cite{Cooper_2015,Adhikary_2025}, error correction \cite{10.1145/3695053.3730991}, protocol design \cite{solani}, non-Gaussian charger of battery \cite{adhikary2026}, etc. We believe this work will contribute well to the study of networks of coupled quantum oscillators by establishing our framework to measure the synchronization phenomena in such systems. Moreover, the methodologies developed here will provide a roadmap for implementing sensing applications in superconducting circuits \cite{PhysRevA.97.013811} and trapped ion platforms \cite{Li2025}, enabling active stabilization and manipulation of synchronized quantum states. Last but not least, this study contributes as a theory-experiment bridge apart from the framework of the vdPo, allowing for further investigation into the potential link between the two figures of merit mentioned here.

\section{Data availability statement}

The data that support the findings of this study are not publicly available. The data are available from the authors upon reasonable request.

% The data that support the findings of this study are openly available in this repository at https \cite{zen}. The repository contains all numerical datasets and code needed to reproduce the results without restriction.

\section{Acknowledgments}
K. A. thankfully acknowledges the hospitality of IMSc for his visit to complete this work. M. R. thank the financial support provided for this research by the DST Anusandhan National Research Foundation, Government of India, through the State University Research Excellence scheme, with reference number SUR/2022/003354. K. M. A. acknowledges the financial support from the Department of Science and Technology (DST), Government of India, through the INSPIRE Fellowship with reference number DST/INSPIRE/03/2025/000257 [IF240166].

\appendix
% \section{Master equation of the driven vdPo}\label{aa}

% \section{Elements of the steady state density matrix in the deep quantum regime}\label{bb}

\section{Computational methodology for synchronization measurement}\label{algo}

In this section, we have expanded the discussion to provide a more rigorous numerical analysis to understand our findings, as stated in figure~\ref{arn_tongue}(a), \ref{arn_tongue}b) and figure~\ref{arn_tongue_g2}. It is essential to realize how our two quantifiers, nonclassical area $\delta$ and second-order correlation function $g^{(2)}(0)$, indicate synchronization characteristics. Before going to synchronization measurement, we numerically solved the Lindblad master equation (\ref{mes}) at steady state in QuTiP  \cite{Johansson2013} with a sufficiently large cutoff Hilbert space dimension to ensure convergence. This provides a steady state density matrix $\rho_{\mathrm{ss}}$, which is allowed to directly analyze the effects of various parameters during the synchronization measure with the aforementioned quantifiers. In principle, $\rho_{\mathrm{ss}}$ is utilized to find out average values $\langle |\cdot|\rangle_{\rho_{\mathrm{ss}}}$ as defined in equation~(\ref{Nonclassical_Area}) and equation~(\ref{second_order_correlation}).

We represent the system in the Fock basis $\{|n\rangle\}$ of the harmonic oscillator and truncate the Hilbert space at a finite cutoff $n_{\max}$. All operators are expressed on this basis, enabling the numerical simulation of the Lindblad equation~(\ref{mes}) and derived quantifiers of synchronization.

\bibliographystyle{iopart-num}
\bibliography{reference}
\end{document}